\documentclass{aa}

\usepackage{graphics}

\newcommand{\kms}          {\mbox{${\rm km~s^{-1}}$}}

\def\n2hp{\mbox{N$_2$H$^+$}}
\def\lesssim{\mathrel{\hbox{\rlap{\hbox{\lower4pt\hbox{$\sim$}}}\hbox{$<$}}}}
\def\gtrsim{\mathrel{\hbox{\rlap{\hbox{\lower4pt\hbox{$\sim$}}}\hbox{$>$}}}}

\begin{document}

\thesaurus{06(13.19.3;08.06.2;09.11.1;09.13.2;09.01.1;02.18.7)}

\title{Modeling the millimeter emission from the Cepheus A young stellar
cluster: Evidence for large scale collapse}
\titlerunning{Large scale collapse in Cepheus A}

\author{S. Bottinelli
\and J. P. Williams}
\institute{Institute for Astronomy, University of Hawai'i,
Honolulu, HI 96822}

\offprints{Sandrine Bottinelli}
\mail{sandrine@ifa.hawaii.edu}

\date{Received / Accepted}

\maketitle

\def\startfigcap{\vspace*{8.0\baselineskip}\bgroup\leftskip 0.45in\rightskip 0.45in}
\def\endfigcap{\par\egroup\vspace*{2.0\baselineskip}}
\def\plotfiddle#1#2#3#4#5#6#7{\centering \leavevmode
\vbox to#2{\rule{0pt}{#2}}
\includegraphics{#1}}

\begin{abstract}
Evidence for a large scale flow of low density gas onto the
Cepheus A young stellar cluster is presented. Observations of
K-band near-infrared and multi-transition CS and \n2hp\ millimeter
line emission are shown in relation to a sub-millimeter map of the
cool dust around the most embedded stars. The near-infrared emission
is offset from the dust peak suggesting a shift in the
location of star formation over the history of the core.
The CS emission is concentrated
toward the core center but \n2hp\ peaks in two main cores offset
from the center, opposite to the chemistry observed in low mass
cores. A starless core with strong CS but weak \n2hp\ emission
is found toward the western edge of the region. The average CS(2--1)
spectrum over the cluster forming core is asymmetrically self-absorbed
suggesting infall. We analyze the large scale dynamics by applying
a one-dimensional radiative transfer code to a model spherical
core with constant temperature and linewidth, and a density
profile measured from an archival $850~\mu$m map of the region.
The best fit model that matches the three CS profiles
requires a low CS abundance in the core and an outer, infalling
envelope with a low density and undepleted CS abundance.
The integrated intensities of the two \n2hp\ lines is well
matched with a constant \n2hp\ abundance. The envelope infall
velocity is tightly constrained by the CS(2--1) asymmetry and
is sub-sonic but the size of the infalling region is poorly
determined. The picture of a high density center with depleted
CS slowly accreting a low density outer envelope with normal
CS abundance suggests that core growth occurs at least partially
by the dissipation of turbulent support on large scales.

\keywords{Radio lines: ISM -- Stars: formation -- ISM: kinematics and
dynamics -- ISM: molecules -- ISM: abundances -- Radiative transfer}
\end{abstract}

\section{Introduction \label{intro}}

Most stars, particularly massive stars, form in groups (e.g.
\cite{carpenter00} 2000).
It is therefore essential to study cluster forming regions in order to
understand more completely the way in which the majority of stars are formed.
Isolated low mass star formation occurs via
the nearly isothermal free-fall collapse of a dense molecular
cloud core, followed by the evolutionary phases
Class 0, I, II and III objects (e.g. \cite{evans99} 1999).
However, the applicability of this paradigm to the formation of
massive stars is debated (\cite{garay+lizano99} 1999):
for example, massive stars begin burning hydrogen
and reach the main sequence while still accreting matter from the
surrounding protostellar envelope and they can also develop strong winds,
both of which will strongly affect the physical conditions, structure
and chemistry of their surroundings.
Due to the shape of the IMF and the fact that they evolve faster,
massive protostars are rarer (and therefore more distant on average)
than low mass protostars. Consequently fewer Class 0 massive protostar
counterparts have been studied in detail.
It is only recently that catalogs of high-mass protostellar
objects have been made (e.g. \cite{sridharan02} 2002).

The molecular cloud core Cepheus A East (hereafter Cep A) is a nearby site
of massive star formation (\cite{sargent77} 1977) located in the Cepheus OB
association
at a distance of 725 pc (\cite{blauuw59} 1959). The far-IR luminosity is
2.4$\times 10^4~L_\odot$ (\cite{evans81} 1981),
corresponding to a small cluster of B stars.
Cep A harbors one of the first molecular bipolar outflow sources discovered
(\cite{rodriguez80} 1980).
Higher spatial resolution CO observations showed the outflow
to be extremely complex, and it was termed quadrupolar
(\cite{torrelles93} 1993).
The fastest components of this outflow are bipolar and oriented
northwest-southeast (\cite{rodriguez80} 1980),
perpendicular to the low velocity CO structure.
The slower and more extended component has been interpreted as the diverting
and redirecting of the main outflow by the interaction with interstellar
high-density gas, seen in NH$_3$ lines by \cite{torrelles93} (1993).
Ultra compact
\ion{H}{ii} regions and a diffuse thermal dust emission source have been
identified from 20~$\mu$m maps and 6 cm low-resolution VLA observations by
\cite{beichman79} (1979). Seven ionized hydrogen complexes lie in ``strings''
that form
a ``Y'' tilted to the east (\cite{hughes+wouterloot84} 1984),
the bifurcation point of which is coincident with the exciting
source of the molecular outflow. A cluster of compact radio sources
have been identified as pre-main-sequence stars by \cite{hughes88} (1988)
due to their variability and the presence of OH and H$_2$O maser emission.
From subsequent ammonia VLA observations, \cite{torrelles93} (1993) proposed
that one of these radio sources, HW2, is a $\sim$10--20~$M_\odot$ protostar.
The larger core surrounding the cluster has
temperature 35 K and mass $200-300 M_\odot$ \cite{ms91} (1991).
On the basis of its protostellar content, high luminosity and
low temperature, and following the bolometric temperature definition
of \cite{chen95} (1995), the Cep A core may be considered a high mass
Class 0 source.

In order to examine the properties of this young, massive cluster
forming region, we obtained near-infrared and millimeter wavelength
multi-transition CS and \n2hp\ data.
The observations are discussed in \S\ref{data}.
We derive the density profile from $850~\mu$m continuum measurements
and fit core-averaged spectra using a 1-D radiative transfer model in
\S\ref{analysis}.
Our results indicate large CS depletion in the central core and
an outer undepleted accreting layer. We discuss these results and
conclude in \S \ref{discussion}.

\section{Observations \label{data}}

\subsection{Millimeter Data}

Observations were made with the 10
antenna Berkeley-Illinois-Maryland array\footnotemark
\footnotetext{The BIMA array is operated with support from the
National Science Foundation under grants AST-9981308 to UC Berkeley,
AST-9981363 to U. Illinois, and AST-9981289 to U. Maryland.}
(BIMA) for two 8 hour tracks in CS(2--1) in April and May 1998
and one 8 hour track in \n2hp(1--0) in May 1998, all in C-array.
A seven field hexagonal mosaic was made with phase center,
$\alpha(2000)=22^{\rm h}56^{\rm m}18\fs 9,
\delta(2000)=62^\circ 01' 42\farcs 6$.
Amplitude and phase were calibrated using 5 minute observations of
2322+509 interleaved with each 25 minute integration on source.
The calibrator flux was $0.66\pm 0.22$~Jy based on observations
of Uranus during the middle of each track.
The correlator was configured with two sets of 256 channels at
a bandwidth of 12.5~MHz (0.15~\kms\ per channel) in each sideband
and a total continuum bandwidth of 800~MHz.
Data reduction was carried out using standard procedures
in the MIRIAD package. The final maps covered a hexagonal region
$\sim 3\farcm 5\times 3\farcm 5$ region at $\sim 9''\times 7''$
resolution.

Complementary single-dish maps of the same lines were made
at the Five College Radio Astronomy Observatory\footnotemark
\footnotetext{FCRAO is supported in part by the
National Science Foundation under grant AST-0100793 and is operated
with permission of the Metropolitan District Commission, Commonwealth
of Massachusetts.}
(FCRAO) 14~m telescope in December 1999 using the SEQUOIA 16 beam
array receiver and the FAAS backend consisting of 15 autocorrelation
spectrometers with 1024 channels set to an effective resolution
of 24~kHz (0.06~\kms). The CS and \n2hp\ lines were observed simultaneously
in frequency switching mode. The pointing and focus were checked every
three hours on nearby SiO maser sources. Third order baselines were
removed from the data and spectra coadded using the CLASS package.
The resolution of the data is $50''$ and the final maps were Nyquist
sampled over $8' \times 8'$ centered on the BIMA phase center.
A single spectrum of C$^{34}$S(2--1) was also taken
toward the map center using the same setup.

The FCRAO data were combined with the BIMA data using maximum entropy
deconvolution
(using a gain for FCRAO at these frequencies of 43.7~Jy~K$^{-1}$).
The resulting maps show the large scale structure observed in the
single-dish map at the $\sim 10''$ resolution of the interferometer map.
These maps have an rms noise of 0.2 and 0.4~K per 0.5~\kms\ channel, for
CS (2--1) and \n2hp\ (1--0) respectively.
All the flux is recovered in the combined map but
features at intermediate scales, in the range $70''-100''$,
may be poorly represented (e.g. \cite{williams03} 2003).

Observations of CS(5--4), CS(7--6) and \n2hp(3--2) were made
at the Heinrich Hertz Telescope\footnotemark\footnotetext{The HHT
is operated by the Submillimeter Telescope Observatory on behalf
of Steward Observatory and the Max-Planck-Institut fuer Radioastronomie.}
(HHT) in November 1999. The data were taken using the SIS-230 and SIS-345
receivers and AOS backend (2048 channels, 48~kHz resolution)
in on-the-fly (OTF) mode.
Pointing and focus were checked using observations of Saturn and Orion IRc2.
The final maps, made from several OTF maps in orthogonal scan directions,
were coadded and first order baselines removed using the CLASS package.
The resolution of the observations is $31'', 22''$, and $27''$
for CS(5--4), CS(7--6) and \n2hp(3--2) respectively and the final
maps covered $3\farcm 5\times 3\farcm 5$, with an rms noise of
0.4~K per 0.5~\kms\ channel.

To compare the near-infrared and millimeter wavelength line data with
the cool dust emission around the most embedded (Class 0 counterpart)
stars in the cluster, we downloaded archival SCUBA observations taken
on the James Clerk Maxwell Telescope in August 1997 and reduced them
using the SURF package.

\begin{figure*}
\resizebox{\hsize}{!}{\includegraphics{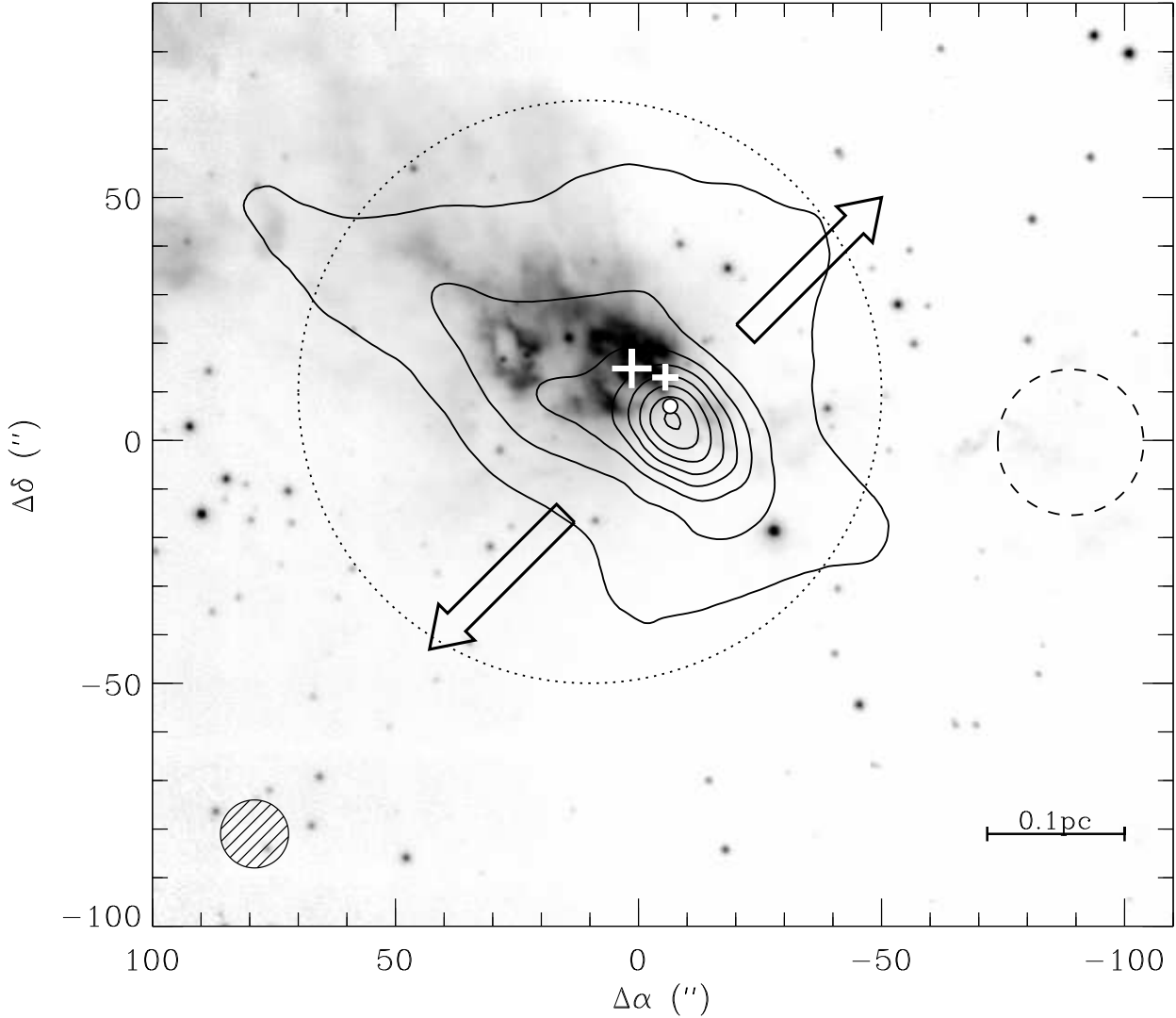}}
\caption{Grayscale K-band image over-plotted with SCUBA 850 $\mu$m
contours of Cep A showing the stellar and dust content.
The central position is at
$\alpha(2000)=22^{\rm h}56^{\rm m}18\fs 9,
\delta(2000)=62^\circ 01' 42\farcs 6$.
The $850~\mu$m contours start at 1 Jy beam$^{-1}$ 
and increment by 2 Jy beam$^{-1}$.
The large and small white plus signs show the location of the
IRAS and MSX point sources respectively
(each with a positional accuracy of less than $5''$).
The small white circle near the center of the map marks the
location of the high mass protostar, HW2. The arrows show the direction
of the bipolar CO outflow observed by \cite{rodriguez80} (1980).
The dashed circle at the western edge of the map shows the location
of the starless core seen in the CS maps
and the dotted circle, centered at (10,10) with diameter $2'$
defines the region over which the spectral line averages were calculated.}
\label{scuba850}
\end{figure*}

\begin{figure*}
\resizebox{\hsize}{!}{\includegraphics{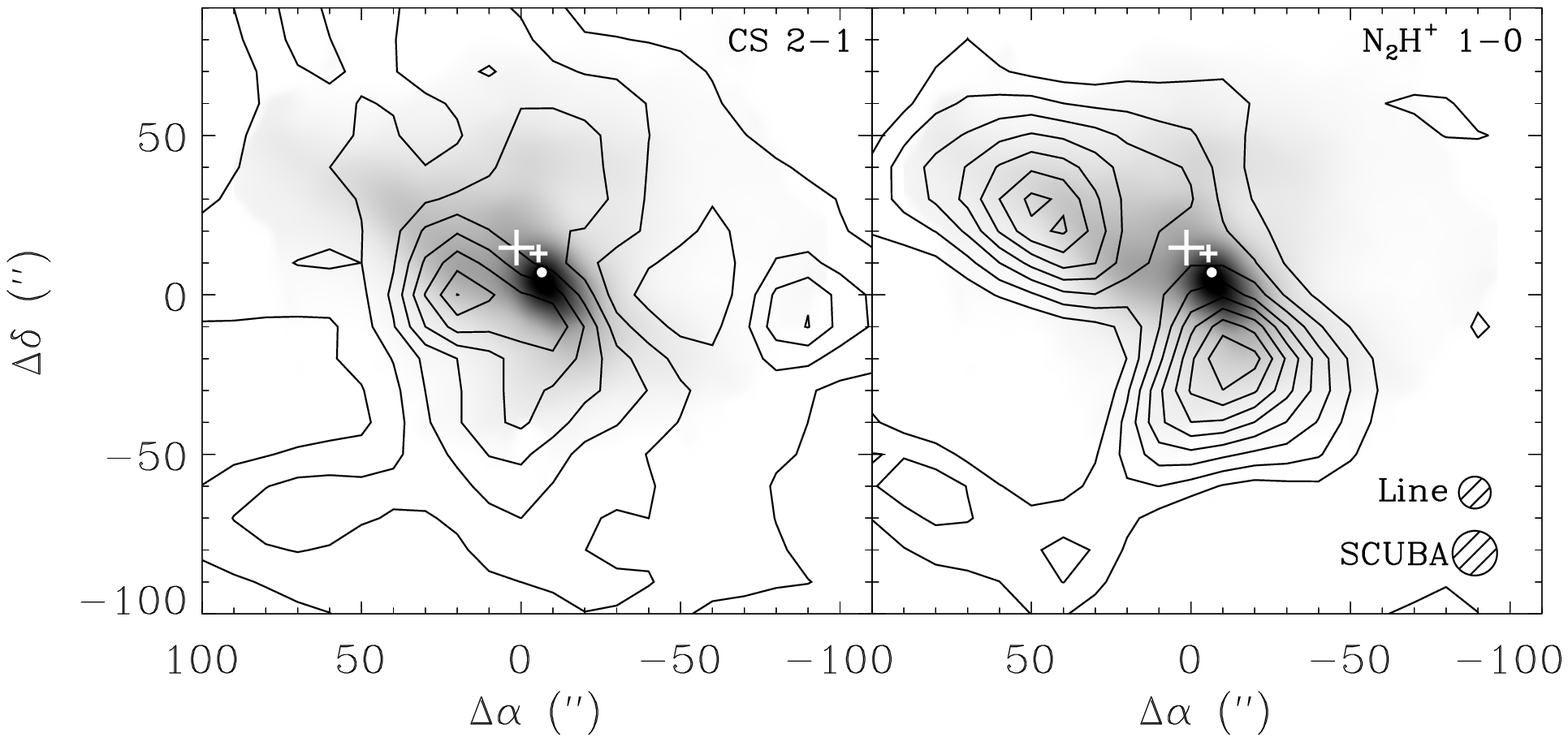}}
\caption{Comparison of line and dust emission.
The line maps have a resolution of $10''$ and are
contoured with starting levels and increments equal to
4.0~K~\kms\ for CS (2--1) and 5.0 for \n2hp (1--0).
The dust emission has a resolution of $14''$ and
is shown as a grayscale ranging from 1 to 15 Jy beam$^{-1}$.
Symbols are as in Figure~\ref{scuba850}.}
\label{dust+line}
\end{figure*}

\subsection{Infrared Data \label{nir}}

Cep A was observed on the University of Hawai'i 2.2m telescope at the f/10
focus with QUIRC (QUick InfraRed Camera) in
K band (2.200~$\mu$m) in July and August 2003.
The plate scale  of the telescope is 0.1886~arcsec~pixel$^{-1}$ and
the field-of-view $193''~\times~193''$.
The total on source integration time was 23.2~minutes. The average
seeing was $1\farcs 1$ FWHM. The observations were carried out
by taking alternate object and sky exposures. All frames were
flat-fielded using a normalized incandescent light dome flat.
Sky frames were obtained
by computing the median of several sky exposures close in time to a
given object frame. The sky frames were subtracted from the object frames.
The individual images were registered and co-added to produce
a single image.
Unfortunately, the conditions were non-photometric but we conservatively
estimate that we should be able to detect to
a limiting magnitude of at least 19,
which is 3.4 magnitudes fainter than 2MASS.

\section{Analysis \label{analysis}}

\subsection{Continuum and Line Maps\label{results}}

Figure \ref{scuba850} shows contours of the 850 $\mu$m continuum
emission overlaid on the grayscale K-band image.
Note that for 35~K dust and the SCUBA beamsize of $14''$,
1 Jy beam$^{-1}$ corresponds to a visual extinction, $A_{\rm V}=24$,
or an extinction at K-band, $A_{\rm K}=2.4$.
The infrared nebulosity lies within the extended sub-millimeter
emission ($A_{\rm K}\sim 5-8$) but is offset from the peak.
There is one IRAS and one MSX point source in the region,
slightly offset from each other.
The IRAS-HiRes $12~\mu$m and MSX $10~\mu$m emission are both
associated with the near-infrared nebulosity, but the $60~\mu$m
IRAS-HiRes image peak is closer to the maximum of the
$850~\mu$m map, which traces the cooler dust.
This suggests a spread in ages and location of star formation in Cep A,
with the youngest protostars more deeply embedded in the core and invisible
at near- and mid-infrared wavelengths.

\begin{figure}
\resizebox{\hsize}{!}{\includegraphics{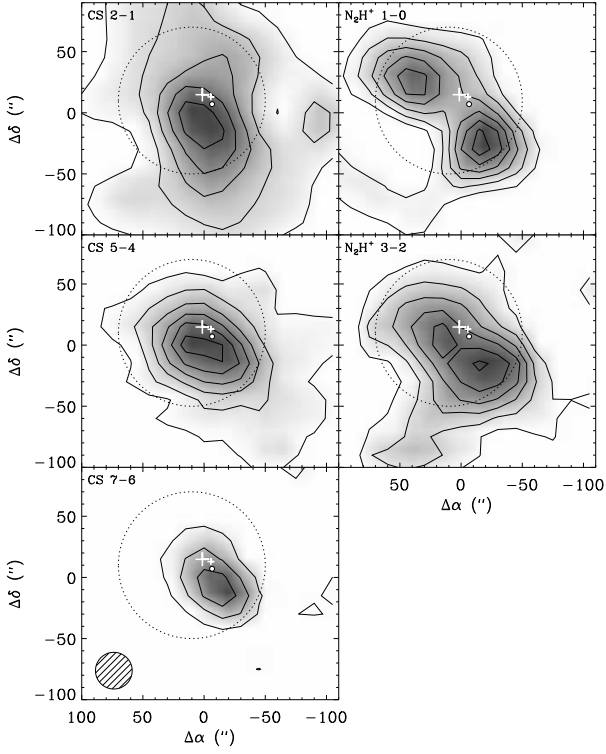}}
\caption{
Line maps of the Cepheus A core shown as grayscale with contours
overlayed. Positions are given in offset
coordinates with the same central position as Figure~\ref{scuba850}.
CS maps are shown in the left panels and \n2hp\ in the right.
The maps have been smoothed to a uniform $30''$ resolution.
The contour starting levels and increments for each map are
4.0~K~\kms\ (CS 2--1), 4.0 (CS 5--4), 1.5 (CS 7--6),
5.0 (\n2hp 1--0), and 3.0 (\n2hp 3--2).
Note the isolated core in the CS(2--1) map
$\sim 90''$ west of center (see text for discussion).
The $2'$ diameter circle shows the region over which the data
were averaged to analyze the large scale dynamics of the core,
and symbols are as in Figure~\ref{scuba850}.}
\label{cs+n2hp}
\end{figure}

Spectral line maps and their comparison with the dust emission
are shown in Figures~\ref{dust+line},~\ref{cs+n2hp}. In each case the
emission has been integrated over the full extent of the line,
including all hyperfine components in the case of \n2hp.
Figure~\ref{dust+line} shows the combined BIMA+FCRAO data at $10''$
resolution. Figure~\ref{cs+n2hp} shows these maps and the higher
transition data smoothed to a uniform resolution of $30''$.
Due to their high dipole moments, CS and \n2hp\ both trace high
volume densities over a range, $n_{\rm H_2}\sim 10^4-10^6$ cm$^{-3}$,
for these transitions.
The complex structure apparent in the line maps contrasts with the
relative simplicity of the dust continuum emission.
The spatial distribution of \n2hp\ is similar to that of
the ammonia (\cite{torrelles93} 1993) and generally follows the
$850~\mu$m continuum emission, but \n2hp\ is notably absent toward
the core center where the CS is strongest.
This behavior is opposite to the situation in low mass cores
where \n2hp\ is more centrally concentrated than CS
(\cite{tafalla02} 2002).

The maps also reveal an apparently starless core located $\sim 90''$
to the west of the bright $850~\mu$m peak, indicated by a cross on
Figure \ref{scuba850}.
The core is detected in CS(2--1) and (5--4) but has only very weak
\n2hp(1--0) emission (Figure~\ref{dust+line})
and no stars are seen in the near-infrared image.
This core is on the edge of the SCUBA map where there is extended
emission but no significant peak. The presence of
CS (5--4) emission indicates large enough densities that we would expect
to observe \n2hp\ emission, as in the large core. Since this emission
is very weak, there must be a variation of abundances between the two cores.
Potential reasons for the wide range of abundances throughout the
region are discussed in \S \ref{discussion}.

\subsection{Radiative Transfer Modeling \label{ratran}}

\begin{figure}
\resizebox{\hsize}{!}{\includegraphics{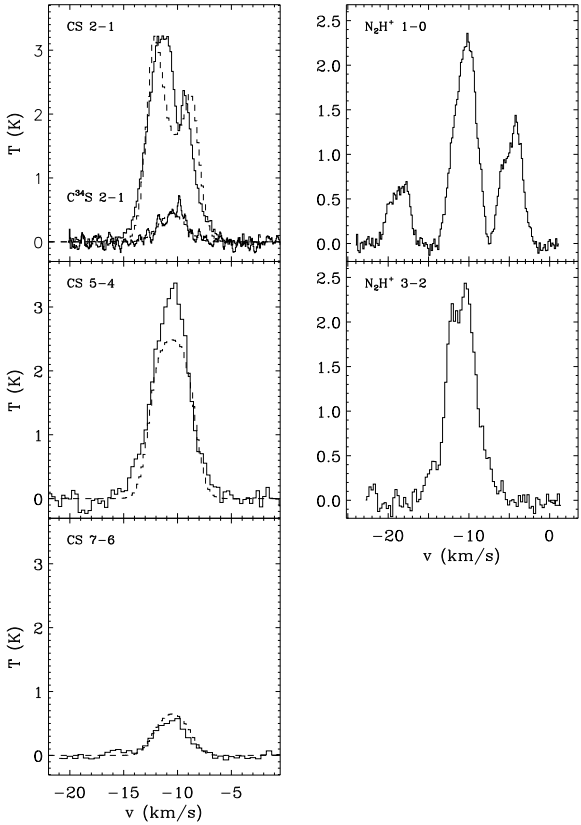}}
\caption{Averages of CS and C$^{34}$S (2--1) (left)
and \n2hp\ (right) spectra over the
central core region outlined by the circles in Figures~1 and \ref{cs+n2hp}.
The heavy solid lines show the observed data and the dashed line
shows the modeled profiles for the CS lines. The multiple peaks
in the \n2hp\ spectra is due to hyperfine structure which was not
modeled (the fits matched only the integrated intensity of each
rotational transition).}
\label{cs_n2hp_spec0}
\end{figure}

Our understanding of the structure and dynamics in Cep A is complicated
by the small scale chemical variations within the core.
Henceforth, we restrict attention to the
large scale properties of the region as if observed with a
$2'$ gaussian beam centered at offset (10,10).
The size and location of the averaging region was chosen so as
to be broadly centered on the sub-millimeter continuum map but
also to include the most prominent CS and \n2hp\ structures within
the core. The averaging region is indicated on the continuum and
line maps in Figures \ref{scuba850} and \ref{cs+n2hp}.
Averaged line profiles are displayed in Figure \ref{cs_n2hp_spec0}.
This figure also includes the line profile for C$^{34}$S(2--1). This
isotopomer of CS, which is less abundant and therefore optically thinner,
peaks near the same velocity as the dip in the average CS(2--1) spectrum.
We therefore conclude that the two peaks in the latter are due to
radiative effects (self-absorption) and not two features along
the line of sight. Although individual CS profiles show both
blue- and red-shifted self-absorption that may be
due to a mix of infall, outflow, and rotation (\cite{difrancesco01} 2001),
the average CS(2--1) profile has a
blue-shifted peak brighter than the red-shifted one, indicating
that infall is the dominant effect on the scale of the core
(\cite{leung+brown77} 1977).

In order to examine the properties of the large scale structure and
dynamics, we modeled the core average line profiles using
the radiative transfer code, ratran (\cite{hvdt00} 2000).
Collisional rate coefficients for CS are from \cite{turner92} (1992)
and for \n2hp\ from \cite{monteiro85} (1985 and references therein).
From the brightness temperature of the \n2hp\, and assuming an 
excitation temperature equal to the dust temperature, 35~K,
we estimate the optical depth of each hyperfine component
to be less than 0.1. For the purposes of the modeling, therefore,
we can ignore the complicated hyperfine structure (7 components at
$J$ = 1--0 and 45 at $J$ = 3--2) and simply match the integrated
intensity of each rotational transition.

The inputs to the radiative transfer model include the density,
temperature, and velocity structure of the core and the abundances
of each molecule. However, many parameters are constrained by
related observations and results. The remaining free parameters,
eight in all, were then varied so as to fit the profiles of each
of the three CS transitions and the integrated intensity of the
two \n2hp\ transitions.

\subsubsection{Model inputs}

The $850~\mu$m map in Figure~\ref{scuba850}
allows us to measure the column
density profile of the core and thereby estimate the volume density.
Since the radiative transfer model is one-dimensional, we approximated
the density structure as a radial function by calculating the average
flux in concentric elliptical annuli centered on the peak of the emission.
The equivalent radii were defined as the geometric mean of the semi-major
and semi-minor axes of each ellipse. The column density was determined
by assuming a dust temperature of 35~K (\cite{ms91} 1991)
and a mass-opacity $\kappa=0.02~{\rm cm}^2~{\rm g}^{-1}$
(\cite{oh94} 1994).
The volume density was then derived by assuming a path length through
the core equal to twice the equivalent radius. The resulting density
profile was fit to a Plummer-like model,
$$n_{\rm H_2}(r) = n_0 \bigg[1+\left({r\over r_0}\right)^2\bigg]^{-\alpha/2},$$
where $n_0$ = 1.1$\times 10^7$ cm$^{-3}$, $r_0$ = 0.02 pc and $\alpha$ = 2.0.
The model discretizes this density profile over nine logarithmically
spaced shells. Both measured and discretized density profiles are
displayed in Fig.~\ref{paramplot}; there are no data points for radii
smaller than 0.03~pc, the equivalent radius corresponding to a
semi-minor axis of half the $850~\mu$m beamsize, and for radii larger
than 0.18~pc, which corresponds to the extent of the $850~\mu$m map.

Additional inputs to the model include a temperature of 35~K,
derived from a graybody fit to the SED (\cite{ms91} 1991),
and a systemic velocity and velocity dispersion
derived from fitting the optically thin \n2hp(1--0) line,
$v_{\rm core}=-10.5$~\kms\ and $\sigma_{\rm core}=1.2$~\kms, respectively.
A constant temperature and velocity dispersion were adequate to
model the core.

To model the observed self-absorption in the CS(2--1) line, we found that
we required a low excitation (and therefore low density)
and low velocity dispersion ($\sigma_{\rm env}=1.1$~\kms) outer shell.
The low velocity dispersion is required by the narrowness of
the observed absorption dip.
The overall model has 10 shells therefore, but it can effectively be
considered as two layers; a power law core on the inside and a low
density outer envelope. The free parameters in the model are the
inner core cutoff radius, the size, density, velocity dispersion
and relative velocity of the envelope and the molecular abundances
in the core and envelope.

The values for the molecular abundances were guided by observations
of other cores and theoretical models.
In cold molecular cloud cores prior to star formation, the chemistry is
dominated by low-temperature gas-phase ion-molecule and neutral-neutral
reactions (\cite{vdb98} 1998).
During the cold collapse phase, however, the density becomes so high
that many molecules freeze onto grain surfaces.
\cite{tafalla98} (1998) find that, in low mass star forming cores,
the abundances of the tightly bound sulfur-bearing molecules
such as CS begin to exhibit large depletions at densities in the range,
$n_{H_2} \sim 2-6\times 10^4$ cm$^{-3}$.
This behavior is in contrast to that of \n2hp\ which, due to
the low binding energy of the precursor molecule N$_2$,
depletes only at the highest densities (\cite{bl97} 1997).
Based on these results, we assume a constant abundance of \n2hp\
but allow the CS abundance to vary. As we show later, we fit
the data with a simple ``jump'' model where the CS is depleted in
the inner dense core relative to a lower density outer envelope.
This is a similar abundance profile to the detailed chemical modeling
of the high mass star forming region, AFGL 2591, by \cite{doty02} (2002).

\subsubsection{Model results}

\begin{figure}
\resizebox{\hsize}{!}{\includegraphics{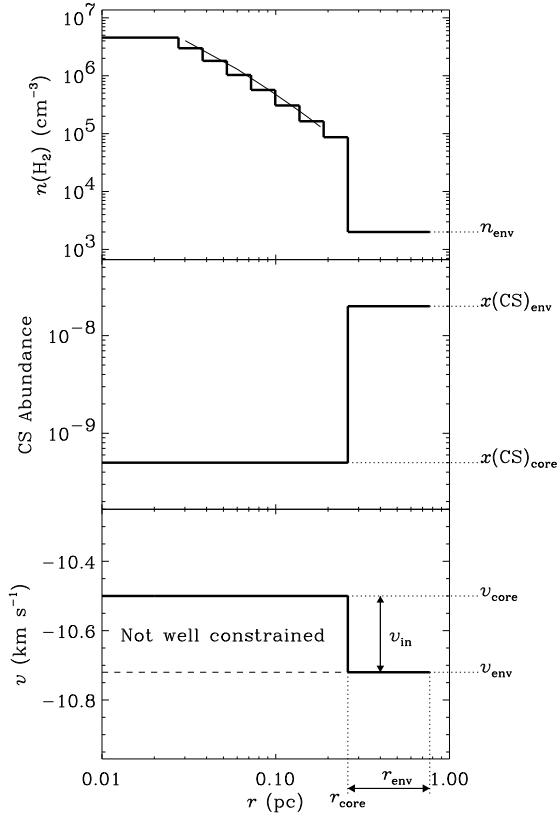}}
\caption{Radial profiles of H$_2$ volume density, core-envelope
CS abundance and velocity profiles used in the best fit model.
Relative motions in the inner region, $r<r_{\rm core}$,
were not well constrained by the data. The thin solid line shows the
H$_2$ density profile derived from the $850~\mu$m data (see text).}
\label{paramplot}
\end{figure}

By varying the density, size and velocity of the envelope, and the CS
abundance in the core and envelope, we were able to fit the integrated
intensities of the three observed CS transitions to within 20\% and
reproduce the asymmetry in the (2--1) line reasonably well. The observed
and model CS spectra are compared in Figure \ref{cs+n2hp}.
The parameters of the fit are tabulated in Table~\ref{param_tbl}, where
$r_{\rm core}$ is the inner core cut-off radius,
$r_{\rm env}$, $n_{\rm env}$ and $v_{\rm in}=v_{\rm env}-v_{\rm core}$
are the size, density and relative velocity of the outer envelope, and
$x$(CS)$_{\rm core}$ and $x$(CS)$_{\rm env}$ are the CS abundances
in the core and the envelope respectively.
The density, velocity, and CS abundance profiles are graphed in
Figure \ref{paramplot} and annotated with the above parameters.
Simultaneously, the \n2hp\ (1--0) and (3--2) integrated intensities were
also matched: we found 8.1 and 9.5 K \kms\ respectively for the
modeled values, whereas the observed values are 9.7 and 8.2 K \kms\,
for the (1--0) and (3--2) transitions respectively.
These results correspond to a match of 16 and 15\% respectively.
Finally, we also fitted the C$^{34}$S (2--1) integrated intensity, 
using an abundance $x$(C$^{34}$S) = $x$(CS)/30. We found a modeled value
of 1.4 K \kms\ which is within 8\% of the observed value (1.5 K \kms).

Despite the good overall fit, some slight discrepancies remain.
The observed CS(5--4) is stronger than the model by 20\%.
This may be due to calibration error or may reflect a more
complex CS abundance profile in the core.
The observed CS(2--1) profile is slightly narrower than the model spectrum,
perhaps due to the constant linewidth assumption in the core
(the velocity dispersion of cores is expected to decrease with radius,
\cite{larson81} 1981).
A more refined model with additional parameters
would fit the data more closely but probably not with greater significance.

Since the model reproduces the blue-red peak asymmetry in the
CS(2--1) spectrum well we are confident that we have accurately
measured the average infall speed of the envelope, $v_{\rm in}=-0.22$~\kms.
However, the size of the infalling region is not well determined
because the average profiles of the higher transition CS lines are
not self-absorbed and therefore relative motions of the higher
density gas could not be constrained.
Nevertheless, we were unable to model the spectra with a static outer shell 
and an inner collapse: no such set of parameters could reproduce the 
dip in the CS (2--1) spectrum. That is, our results indicate a
{\it large scale collapse from the outside-in.}

\begin{table}
\caption[]{Fit parameters.\label{param_tbl}}
\begin{center}
\begin{tabular}[t]{cc}
\hline
\hline
Parameter & Value\\
\hline
$r_{\rm core}$               & 0.26 pc\\
$r_{\rm env}$                & 0.51 pc\\
$n_{\rm env}$                & 2$\times$10$^3$ cm$^{-3}$\\
$\sigma_{\rm env}$           & 1.1 \kms\\
$v_{\rm in}$                 & $-0.22$ \kms\\
$x$(CS)$_{\rm core}$         & $5\times$10$^{-10}$\\
$x$(CS)$_{\rm env}$          & $2\times$10$^{-8}$\\
$x({\rm N_2H^+})$            & $3\times$10$^{-11}$\\
\hline
\end{tabular}
\end{center}
\end{table}

\section{Discussion and Summary \label{discussion}}

The line observations reveal a complex chemistry in the core.
The CS emission is concentrated toward the center near the
peak of the $850~\mu$m dust emission and the youngest, most
embedded protostars. However, the \n2hp\ map shows two
prominent cores offset on either side of
the dust peak. The presence of CS and absence of \n2hp\ toward
the star forming center of the core is likely due to the fact
that neutral molecules released from the dust grains
in the hotter region surrounding the protostars preferentially
destroy ions such as \n2hp\ (\cite{bergin00} 2000).

The maps also reveal a small starless core toward the west of the main
core. The relatively strong CS and weak \n2hp\ emission toward this
core suggests that it has only recently formed (\cite{bl97} 1997).
\cite{williams+myers99} (1999) found a starless core with
similar chemical properties in the Serpens NW cluster

Despite the chemical complexities on small scales, the average
CS(2--1) spectrum is asymmetrically self-absorbed suggesting
large scale collapse. Using the Hogerheijde \& van der Tak
radiative transfer code, we fit the average CS and \n2hp\ spectra with
a spherical model consisting of an inner region with a Plummer-like
density profile measured from archival SCUBA $850~\mu$m data with
constant temperature and linewidth.
The best fit model that matches the three CS profiles has a low
CS abundance in the inner region and an outer, infalling envelope
with a low density and higher CS abundance. The depletion toward
the center matches chemical evolution model expectations
(\cite{bl97} 1997) and the envelope CS abundance is similar to
that in the extended ridge of Orion (\cite{vdb98} 1998).
The fit also matched the integrated intensities of the two \n2hp\
spectra with a constant abundance similar to that found in B68
(\cite{bergin02} 2002). In practice the maps show that the \n2hp\
must deplete toward the core center (see also \cite{doty02} 2002)
but we have not attempted to match the complicated small scale
chemistry and our model fit for the \n2hp\ abundance should only
be considered an average, weighted by column density, over the core.

The CS(2--1) self-absorption requires a large scale
outside-in collapse. The velocity of the collapse could be accurately
measured but the depth of the collapse region could not, due to the
absence of self-absorption in the higher transition CS lines.
At 0.22~\kms, the infall velocity is sub-sonic for a gas temperature
of 35~K. The mass infall rate of the envelope can be estimated from the
ratio of its mass, $M_{\rm env}=240~M_\odot$ determined from its size
and density, and the time for the outer edge to reach the core center,
$$\dot M_{\rm in} = {M_{\rm env}v_{\rm in}\over r_{\rm env}+r_{\rm core}}
                  = 7\times 10^{-5}~M_\odot {\rm yr}^{-1}.$$
This may only be a lower limit to the total mass infall rate at the
center if the inner region is also collapsing. Nevertheless, the
envelope mass infall rate alone is more than an order of magnitude higher than
typical mass infall rates for solar mass protostars (\cite{zhou95} 1995).

Our data do not rule out inside-out collapse motions around individual
protostars at higher densities on smaller size scales since
the $\tau=1$ surface of the CS(2--1) emission occurs at low densities and
therefore at large scales. The multitude of sources, powerful outflows,
and the complex chemistry would likely make an investigation of the
small scale motions around individual protostars quite challenging.

On large scales, however, our picture is of a core with a Plummer-like
density profile accreting low density gas sub-sonically.
The CS abundance in the infalling envelope is similar to
undepleted values in the ISM. The dissipation of turbulent support
resulting in ``cooling flows'' may lead to core growth in this manner
(\cite{nakano98} 1998 ; \cite{myers+lazarian98} 1998 ; 
\cite{williams+myers00} 2000).
The growing availability of dust continuum maps and multi-transition,
multi-species line observations will lead to more refined structural
and dynamical modeling and comparisons between different star
forming environments in the future.

\acknowledgements
This work began when JPW was a Jansky fellow and he thanks the NRAO
for their support. The more recent analysis was supported by
NSF grant AST-0324328.
We thank Harold Butner for assistance with the observations at the
Heinrich Hertz Telescope, Ted Bergin, Eric Herbst and Phil Myers for
informative discussions, and the referee, Gary Fuller, for useful
suggestions that improved the paper.

{}

\end{document}